\documentclass[twocolumn,showpacs,preprintnumbers,amsmath,amssymb]{revtex4}
\usepackage{graphicx}% Include figure files
\usepackage{dcolumn}% Align table columns on decimal point
\usepackage{bm}% bold math

\def\dslash{/\kern-.1em \partial}
\def\trace{\hbox{Tr}}
\def\ldet{\hbox{logdet}}
\def\deter{\hbox{det}}

\def\al{\alpha}
\def\be{\beta}

\def\de{\delta}
\def\ep{\epsilon}

\def\ka{\kappa}
\def\la{\lambda}

\def\Ga{\Gamma}

\def\La{\Lambda}

\def\mn{{\mu\nu}}

\def\frac#1#2{{\textstyle{{#1}\over {#2}}}}

\def\lsim{\mathrel{\rlap{\lower4pt\hbox{\hskip1pt$\sim$}}
    \raise1pt\hbox{$<$}}}
\def\gsim{\mathrel{\rlap{\lower4pt\hbox{\hskip1pt$\sim$}}
    \raise1pt\hbox{$>$}}}
\def\sqr#1#2{{\vcenter{\vbox{\hrule height.#2pt
         \hbox{\vrule width.#2pt height#1pt \kern#1pt
         \vrule width.#2pt}
         \hrule height.#2pt}}}}

\newcommand{\beq}{\begin{equation}}
\newcommand{\eeq}{\end{equation}}
\newcommand{\bea}{\begin{eqnarray}}
\newcommand{\eea}{\end{eqnarray}}

\begin{document}

\preprint{NCF/002}

\title{One-Loop Renormalization of QCD with Lorentz Violation}% Force line breaks with \\

\author{Don Colladay}
\email{colladay@ncf.edu}
\author{Patrick McDonald}%
 \email{mcdonald@ncf.edu}
\affiliation{%
New College of Florida\\ Sarasota, FL, 34243, U.S.A.
}%

\date{\today}% It is always \today, today,

\begin{abstract}
The explicit one-loop renormalizability of the gluon sector of QCD with Lorentz violation
is demonstrated.  The result is consistent with multiplicative
renormalization as the required counter terms are consistent with a
single re-scaling of the Lorentz-violation parameters. In addition,
the resulting beta functions indicate that the CPT-even
Lorentz-violating terms increase with energy scale in opposition to
the asymptotically free gauge coupling and CPT-odd terms. The
calculations are performed at lowest-order in the Lorentz-violating
terms as they are assumed small.

\end{abstract}

\pacs{Valid PACS appear here}% PACS, the Physics and Astronomy
\maketitle

\section{\label{sec:intro}Introduction}

As is well known, the Standard Model is defined by a Lagrangian which
exhibits ultraviolet divergences arising from the structure of the
theory at small distances.  These divergences can be removed by a
singular redefinition of the parameters defining the theory through
the process of renormalization.  The required calculations involve a
number of remarkable cancellations which are most easily obtained by
exploiting the various symmetries of the theory.  The symmetries
of the conventional Standard Model include the Lorentz group.

The investigation of the renormalizability properties of the SM in the
presence of Lorentz violation began in \cite{klp1,klp2} where the authors
studied one-loop radiative correction for QED with Lorentz
violation.  These calculations were carried out in the framework of an
explicit theory called the Standard Model Extension (SME), which has been
formulated to include possible Lorentz-violating background couplings
to Standard Model fields \cite{ck1,ck2}. In this theory, the one-loop
renormalizability of general Lorentz and CPT violating QED has been established
\cite{klp1}.  
The manuscript \cite{klp1} includes an analysis of the explicit one-loop
structure of Lorentz-violating QED and the resulting running of the
couplings.  The authors establish that conventional multiplicative
renormalization succeeds and they find that the beta functions indicate a variety of
running behaviors, all controlled by the running of the charge.
Portions of this analysis have been extended to allow for a
curved-space background \cite{bps}, while other analysis involved
finite, but undetermined radiative corrections due to CPT violation
\cite{kafstuff1,kafstuff2,kafstuff3,kafstuff4,kafstuff5,kafstuff6,kafstuff7}.

The Lorentz violating QED results of \cite{klp1} were extended to non-abelian gauge
theories in \cite{cm1} where the authors established that the associated pure
Yang-Mills theory is renormalizable at one-loop.  More precisely,
conventional multiplicative renormalization succeeds  and the beta
functions indicate that CPT-even Lorentz violating terms increase with
energy scales in opposition to the asymptotically free gauge couplings
and the CPT-odd couplings.  The primary purpose of this paper is to
extend the results of \cite{cm1} to include fermions as well as to remove
certain technical restrictions on the trace of the CPT-even terms.
The results are directly applicable to
one-loop renormalizability of the gluon sector of QCD in the presence
of Lorentz violation.
In addition, the running of the couplings are studied using the
associated beta functions.  New methods for
carrying out renormalization calculations inside the SME
using functional determinants are provided, giving a
second, alternative derivation of the results presented in \cite{cm1} .

This work should be viewed as part of an extensive, systematic
investigation of Lorentz violation and its possible implications for
Planck-scale physics \cite{kps1,kps2,kps3,kps4,kps5,kps6,kps7,kps8,kps9}.
Extensive calculations using the SME
have led to numerous experiments (see, for example \cite{cpt04}),
which have in turn
placed stringent bounds on parameters in the theory associated with
electrons, photons, neutrinos, and hadrons. Recent work involving
Lorentz violation and cosmic microwave background data \cite{kmewes} suggest that
the SME might play a useful role in cosmology.  In
addition to the above, the SME formalism has been extended to include
gravity \cite{alangrav1,alangrav2,alangrav3}, where it has been suggested that Lorentz
violation provides an alternative means of generating General
Relativity \cite{kp}.

Some other related work includes a study of deformed instantons in the theory \cite{cmjmp1,cmjmp2},
an analysis of the Casimir effect in the presence of Lorentz violation
\cite{ft}, an analysis of gauge invariance of Lorentz-violating
QED at higher-orders \cite{ba}, and possible effects due to nonpolynomial interactions \cite{kalt}.
Some investigations into possible
Lorentz-violation induced from the ghost sector of
scalar QED have also been performed \cite{altghost}.
More recently, functional determinants have been used to compute finite corrections
to CPT-violating gauge terms arising from fermion violation \cite{funcdet}.

\section{\label{sec:nandc}Notation and Conventions}

To simplify notation we limit our investigation to the case of a
single fermion.  The associated  Lagrangian with Lorentz
violation is taken to be
\beq
{\mathcal L} =  {\mathcal L}_{A} +  {\mathcal L}_{\psi} + {\mathcal
 L}_{G}~,
\eeq
where ${\mathcal L}_{A}$ is the gauge field contribution, ${\mathcal
  L}_{\psi}$ is the fermionic contribution, and ${\mathcal L}_{G}$
  is the ghost contribution.  In computing the UV divergence, we treat
  each term in the Lagrangian separately.  We begin with the pure
  Yang-Mills contribution \cite{ck1,ck2}
\begin{multline}\label{fl2.1}
{\mathcal L}_{A}  =   - {1 \over 2} tr \left[ F^\mn F_\mn +  (k_F)_{\mn \alpha \beta} F^\mn
  F^{\al \be} + \right. \\
  \left.
 (k_{AF})^\kappa \epsilon_{\kappa \lambda \mu \nu}
  (A^\lambda F^{\mu \nu}   - \frac{2}{3}igA^\lambda A^\mu A^\nu)
+ 2 \lambda {\mathcal F}[A]^2 \right]~,
\end{multline}
where $(k_F)_{\mn \alpha \beta}$ and $(k_{AF})^\kappa$ are tensors
governing the Lorentz violation in the Yang-Mills sector.  For the
Standard Model Extension the tensor $k_F$ is CPT-even, satisfies a
  Jacobi identity and is constrained to have the symmetries of the Riemann
tensor, while the tensor $k_{AF}$ is CPT-odd.
The parameter $\lambda$ multiplies a gauge fixing term $\mathcal F$.
The generators of the
  Lie Algebra defined by $A^\mu = A^{a \mu} t^a$ are taken to satisfy
\beq
[t^a, t^b] = i f^{abc} t^c ~,
\eeq
and $f^{abc}$ are totally anti-symmetric structure constants.
The product of these generators is normalized to
\beq
tr[t^a t^b] = C(r) \de^{ab}~,
\eeq
where $C(r)$ depends on the representation $r$.
In the adjoint representation used for the gauge fields, this is written
$C(G) = C_2(G)$ where $C_2(r)$ is the quadratic Casimir operator
\beq
t^a t^a = C_2(r)\cdot {\bf 1}~.
\eeq
The field tensor is defined as
\beq
F^\mn = - {i \over g}[D^\mu,D^\nu]~,
\eeq
where the covariant derivative is $D^\mu = \partial^\mu + i g A^\mu$.

The fermionic contribution contribution to the Lagrangian is given by
\cite{ck1,ck2}
\beq
{\mathcal L}_{\psi} =  \bar{\psi} ( i\Gamma^\mu D_\mu - M)\psi~,
\eeq
where $ \Gamma^\nu = \gamma^\nu +\Gamma^\nu_1,$  $M=m+M_1,$
 and $\Gamma_1$ and $M_1$ are of the form
\bea
\Gamma_1^\nu & = & c_{\nu\mu}\gamma^\mu + d_{\mu\nu}\gamma_5 \gamma_\mu
+ e_\nu + if_{\nu}\gamma_5 + \frac12 g^{\lambda \mu \nu}
\sigma_{\lambda \mu} \label{gamma2.1} \\
M_1 & = & a_\mu\gamma^\mu + b_\mu\gamma_5 \gamma^\mu + \frac12
H_{\mu\nu}\sigma^{\mu\nu}\label{mass2.1}.
\eea
Here the $\gamma_\mu$ are the standard gamma matrices,
$\sigma_{\lambda \mu}$ are the standard sigma matrices, and the
remaining small parameters control Lorentz violation.  The parameters
$c_{\nu\mu}$ and $d_{\nu\mu}$ are traceless, $H_{\mu\nu}$ is
  antisymmetric, and $g^{\lambda\mu\nu}$ is
antisymmetric in the first two components.  The parameters $a_{\mu}, \
b_\mu, \ H_{\mu\nu},$ have the dimension of mass, while the remaining parameters are
dimensionless.

Finally, the ghost Lagrangian is written in terms of the scalar,
anticommuting field $\phi$
\beq
{\mathcal L}_{G} = - \overline \phi ({\mathcal M}
- C_{\mu\nu}D^\mu D^\nu) \phi ~,
\eeq
where ${\mathcal M}$ is the variation of the gauge fixing functional ${\mathcal F}$
with respect to the gauge transformation
and the constants $C_{\mu\nu}$ parameterize possible Lorentz
violation in the ghost sector \cite{altghost}.

\section{\label{sec:func}Functional Determinants and Background Fields}

Recall, the one-loop effective action for the theory can be written as a functional
integral over fields $\Psi$:
\beq
\exp{i\Ga[\Psi]} = \int{\mathcal D}\Psi e^{i\int d^4 x {\mathcal
    L}[\Psi]}~.
\eeq
The effective action is constructed by writing the underlying fields as the sum
of a classical background and a fluctuating quantum field.
The effective action is given by
a classical term perturbed by terms quadratic in the fluctuation.  The
quadratic term gives rise to a Gaussian integral, which in turn can
be described by a functional determinant \cite{ps}.  Using
${\mathcal L}_{cl} = {\mathcal L}_0 + {\mathcal L}_{c.t.}$
for the classical Lagrangian as a function of
the background field where ${\mathcal L}_{c.t.}$  is the counterterm Lagrangian,
the expression becomes
\begin{equation}\label{expldet3.1}
\exp{i\Ga[\Psi]} = e^{i\int d^4x {\mathcal L}_{cl}}
   \deter(\Delta_A)^{-\frac{1}{2}}
  \deter(\Delta_\psi)^{\frac{1}{2}} \deter(\Delta_\phi)~,
\end{equation}
where the $\Delta$ are operators which are given explicitly below.  To
compute the above determinants, dimensional regularization is used.
Each determinant is treated separately, beginning with the pure Yang-Mills
gauge field contribution.  The calculation is performed to first order in Lorentz violating
parameters.  As this is the case, the computations of the various terms decouple
and the CPT-even and CPT-odd cases can be treated independently.

We will write the gauge fields as the sum of a classical background
field (denoted with an underline) and a fluctuating quantum field:
\bea
A^\mu & = & \underline{A}^\mu +{\mathcal A}^\mu.
\eea
With this convention the curvature can be expressed as
\bea
F^{a \mu\nu} =& & \underline{F}^{a\mu\nu} + \left(\underline{D}^\mu {\mathcal
  A}^\nu\right)^a - \left(\underline{D}^\nu {\mathcal A}^\mu\right)^a \nonumber \\ & & -
gf^{abc}{\mathcal A}^{b\mu}{\mathcal A}^{c\nu} ~, \label{curv2.1}
\eea
where the underline denotes background curvature and the covariant
derivatives are taken with respect to the background fields.
The gauge fixing functional is chosen to be
\beq
{\mathcal F}[A] = \underline D^\mu {\mathcal A}_\mu~,
\eeq
and $\lambda$ is set equal to 1 incorporating Feynman gauge.
Rescaling the vector potential to absorb $g$, substituting (\ref{curv2.1}) into
(\ref{fl2.1}) and retaining terms which are quadratic in the
perturbation of the background field we have
\begin{multline}\label{quad2.1}
{\mathcal L}_{{\mathcal A}}^{quad} =  -\frac{1}{2g^2} tr {\mathcal A}^{\mu} \left[
    -g_{\mu\nu}\underline{D}^2-2i\underline{F}_{\mu\nu}\right.  \\  \left.
   - i (k_F)_{\alpha\beta\mu\nu} \underline{F}^{\alpha\beta}
    -2(k_F)_{\mu\alpha \nu\beta}\underline{D}^\alpha \underline{D}^\beta \right.\\ \left.
    - (k_{AF})^\kappa \ep_{\ka\la\mn} \underline D^\la
    \right]{\mathcal A}^{\nu}~.
\end{multline}
The trace is performed over the Lie Algebra indices and all fields are written
in the adjoint representation.
The trace is extended to cover the Lorentz indices as well using
the matrix notations
\bea
(\tau_{\alpha \beta})_{\mu\nu} & = & i(g_{\alpha\mu}g_{\beta\nu} -
g_{\alpha\nu}g_{\beta\mu}) \\
(\ep_{\al\be})_\mn & = & \ep_{\al\be\mn} \\
(k_{F\alpha \beta}^I)_{\mu\nu} & = &(k_F)_{\alpha\beta\mu\nu} \\
(k_{F\alpha \beta}^{II})_{\mu\nu} & = &(k_F)_{\mu\alpha\nu \beta}~,
\eea
the quadratic Lagrangian can be rewritten as
\bea
{\mathcal L}_{{\mathcal A}}^{quad} & = & -\frac{1}{2g^2}tr {\mathcal A}
  \left[ (-g_{\alpha\beta} -2 k_{F\alpha
      \beta}^{II})\underline{D}^\alpha \underline{D}^\beta
      \right. \nonumber \\ & & \left.
    - \left( \tau_{\alpha \beta} + i
      k_{F\alpha \beta}^I \right)\underline{F}^{\alpha\beta} \nonumber
      - (k_{AF})^\al \ep_{\al\be} \underline D^\be \right]{\mathcal A} \nonumber \\
 &  = & -\frac{1}{2g^2} tr {\mathcal A} \Delta_A {\mathcal A}~,
\eea
where the trace is performed over both Lorentz and gauge spaces,
$\Delta_A = P_A+\Delta_A^{(1)} + \Delta_A^{(2)} + \Delta_A^{(F)},$
$\Delta_A^{(i)}$ is order $i$ in the fields ($i=1,\ 2$), and
$\Delta_A^{(F)}$ contains all curvature contributions:
\bea
P_A & = & -(g_{\alpha\beta} + 2 k_{F\alpha
      \beta}^{II})\partial^\alpha\partial^\beta - k_{AF}^\al \ep_{\al\be}\partial^\be \nonumber \\
\Delta_A^{(1)} &= &  -i(g_{\alpha\beta} + 2 k_{F\alpha
      \beta}^{II})(\partial^\alpha{\underline A}^\beta + {\underline
  A}^\alpha\partial^\beta) - i k_{AF}^\al \ep_{\al\be}\underline A^\be\nonumber \\
\Delta_A^{(2)} &= &  (g_{\alpha\beta} + 2 k_{F\alpha
      \beta}^{I})({\underline A}^\alpha{\underline  A}^\beta) \nonumber \\
 \Delta_A^{(F)} & = & -[ \tau_{\alpha \beta} + i
      k_{F\alpha \beta}^I] \underline{F}^{\alpha\beta}~.
\eea
We compute an expansion for $\ldet(\Delta_A)$ retaining terms which are
linear in the small parameters $k_F$ and $k_{AF}$.
\beq
\ldet( {P_A}^{-1} \Delta_A)  =  \ldet\left[1+ P_A^{-1}(\Delta_A^{(1)} +
\Delta_A^{(2)} + \Delta_A^{(F)})\right] .
\eeq
Note that $P_A$ is actually independent of the backgroud field
and will cancel out of the functional determinant with proper
normalization.
The determinant is written in terms of the logarithm function
using the relation $\ldet S = \trace \log S$.
The logarithm is then expanded as
\begin{multline}\label{ldet2.22}
\trace \log (P_A^{-1} \Delta_A)  =
 \trace(P_A^{-1}(\Delta_A^{(1)} + \Delta_A^{(2)} + \Delta_A^{(F)}))   \\
- \frac12
\trace((P_A^{-1}(\Delta_A^{(1)} + \Delta_A^{(2)} + \Delta_A^{(F)}))^2) + \hbox{
  h.o.}
\end{multline}
The analysis of the first term appearing in the expansion is
straightforward:  Since the Lie algebra elements trace to zero, the
first term reduces to a quadratic divergence:
\begin{multline}
\trace(P_A^{-1}(\Delta_A^{(1)} + \Delta_A^{(2)} + \Delta_A^{(F)})) =
 \trace(P_A^{-1}\Delta_A^{(2)}) \\
  =  tr \int\frac{d^4k}{(2\pi)^4} \frac{d^4p}{(2\pi)^4} \frac{1}{p^2} (g^{\mu\nu} +
2(k_F^{II})_{\mu\nu}) {\underline A}^{\mu}(k) {\underline A}^{ \nu}(-k)~.
\end{multline}
where, as before, the trace refers to both
Lorentz and gauge space.
To calculate the contribution which arises from the second order terms
in the expansion, note that trace considerations immediately reduce the
problem to studying the contribution arising from terms of the form
$P_A^{-1}\Delta_A^{(1)}P_A^{-1}\Delta_A^{(1)}$ and
$P_A^{-1}\Delta_A^{(F)}P_A^{-1}\Delta_A^{(F)}.$  The first of the above terms
produces a quadratic divergence that exactly cancels the quadratic
divergence arising from the first order terms.  A lengthy computation
employing dimensional regularization gives the total Lorentz-violating
divergent contribution as
\begin{multline}\label{YMcont1.1}
\ldet(P_A^{-1}\Delta_A)  = \frac{i}{(4\pi)^2}  \Gamma(2-\frac{d}{2}) tr
 \int\frac{d^4k}{(2\pi)^4} \\
 \left[ \frac{7}{3}
k_{F \mu \la \nu}^{~~~~~~ \lambda}(k^\mu k^\nu{\underline A}^2 \right.
 \left. -2 k^\mu {\underline A}^\nu k\cdot {\underline A} + k^2 {\underline
 A}^\mu{\underline A}^\nu)  \right. \\
\left. -  (12) k_{F \mu\alpha\nu\beta}(k^\alpha k^\beta {\underline A}^\mu{\underline
  A}^\nu)\right].
\end{multline}
Note that the trace over Lorentz indices has been performed in the above
expression and only the gauge space trace remains.
The contribution from the Lorentz violating CPT-odd terms is finite: there
is no corresponding UV divergence.  This calculation confirms the
results obtained in \cite{cm1}.

We can analyze the contribution arising due to ${\mathcal L}_\psi$ in
a similar manner.  The mass-term $M$ does not contribute any divergences
to lowest order, so it is omitted from the calculation.  The contribution
from the kinetic piece is squared to facilitate computation
\beq
-( \Gamma^\mu D_\mu)^2  =  -P_\psi(1 - P_\psi^{-1}(\Delta_\psi^{(1)} +
\Delta_\psi^{(2)} + \Delta_\psi^{(F)}))~,
\eeq
where
\bea
P_\psi & = & (g^{\alpha\beta} +
\{\gamma^\alpha, \Gamma_1^{\beta}\})\partial_\alpha \partial_\beta \label{p3.1} \\
\Delta_\psi^{(1)} & = &  -i(g^{\alpha\beta} + \{\gamma^\alpha ,
\Gamma_1^\beta\})(\partial_\alpha{\underline A}_\beta + {\underline A}_\alpha\partial_\beta) \\
\Delta_\psi^{(2)} &= &  (g^{\alpha\beta} + \{\gamma^\alpha ,
\Gamma_1^\beta\}){\underline A}_\alpha{\underline A}_\beta\\
 \Delta_\psi^{(F)} & = & -( S^{\alpha \beta} +
   \frac12 [\gamma^\alpha, \Gamma_1^\beta])\underline{F}_{\alpha\beta}
   \label{p3.2},
\eea
where $\gamma$ are the standard gamma-matrices, $\Gamma_1$ is as defined
in (\ref{gamma2.1}), and $S_{\alpha \beta}$ are the generators for the
Lorentz transformations in the spinor representation.  With these
adjustments, the expansion (\ref{ldet2.22}) holds for the case of
${\mathcal L}_\psi.$  Symmetry considerations imply that, to
lowest order in the parameters, terms involving parameters other than
$c_{\mu\nu}$ and $b_\mu$ do not contribute to the UV divergences in
the pure Yang-Mills sector.  Similarly, explicit calculation confirms
that terms involving $b_\mu$ do not contribute to UV divergences.
Thus, for the purpose of
computing UV divergences arising from the Lorentz violating fermionic
Lagrangian given by (\ref{gamma2.1}) and (\ref{mass2.1}), only the
terms involving $c_{\mu\nu}$ are relevant, and we can proceed with the
computation assuming that $\Gamma_1^\mu$ is replaced by
$c^{\mu\nu}\gamma_\nu$ and $M_1$ is replaced by $0$ in
(\ref{gamma2.1}) and (\ref{mass2.1}), respectively, and that the
associated simplifications are made in (\ref{p3.1})-(\ref{p3.2}).
With these simplifications, we proceed as we did in the case of pure
Yang-Mills.  The analysis of the first order contribution leads to a
quadratic divergence.  More precisely, since $\Delta_\psi^{(1)}$ is linear
in the fields and the $\Delta_\psi^{(F)}$ term will trace to zero, we have
the Lorentz-violating contribution
\begin{multline}
\trace (P_\psi^{-1}(\Delta_\psi^{(1)} +\Delta_\psi^{(2)} + \Delta_\psi^{(F)}))  =  \\
tr \int\frac{d^4k}{(2\pi)^4}  \frac{d^4p}{(2\pi)^4} \frac{1}{p^2} (2 c_{\mu\nu}) {\underline
  A}^\mu(k) {\underline A}^\nu(-k).
\end{multline}
Using dimensional regularization and following the same calculation as
was carried out for pure Yang-Mills, the quadratic piece of the divergence is
exactly cancelled by a term arising from a second order contribution.
The total contribution associated to the second term in the expansion
(\ref{ldet2.22})is
\begin{equation}\label{fermcont3.1}
\ldet(P_\psi^{-1} \Delta_\psi)  = -\frac{i}{3}
  \frac{C(r)}{(4\pi)^2} \Gamma(2-\frac{d}{2}) c_{\mu\nu}  Q^{\mu \nu}.
\end{equation}
where $C(r)$ is given by the relation $tr (t_r^a t_r^b) = C(r) \delta^{ab}$
and $r$ refers to the fermion representation.
The other new factor is defined as
\begin{equation}\label{qmunu}
Q^{\mu\nu} = \int\frac{d^4k}{(2\pi)^4} (k^\mu k^\nu{\underline
  A}^2 -2 k^\mu \underline A^\nu k\cdot {\underline A} + k^2 {\underline
  A}^\mu{\underline A}^\nu)~.
\end{equation}
To treat the ghost we write the quadratic contribution
\beq
{\mathcal L}_{G} = - \overline \phi (-\underline D^\mu \underline D_\mu
- C_{\mu\nu} \underline D^\mu \underline D^\nu) \phi ~,
\eeq
and express the functional determinant using
\bea
P_\phi & = & (-\partial^2 - C_{\mu\nu}\partial^\mu\partial^\nu) \\
\Delta_\phi^{(1)} & = &  -i(g_{\mu\nu} + C_{\mu\nu})( {\underline
  A}^\mu\partial^\nu+ \partial^\mu{\underline A}^\nu) \\
\Delta_\phi^{(2)} &= &  (g_{\mu\nu} + C_{\mu\nu})({\underline A}^\mu{\underline  A}^\nu).
\eea
Computing as above and using the notation introduced in \eqref{qmunu},
the ghost contribution is given by
\begin{equation}\label{ghostcont3.1}
\ldet(P_\phi^{-1}\Delta_\phi) =
-\frac{i}{6}\frac {C_2(G)}{(4\pi)^2} \Gamma(2-\frac{d}{2}) C_{\mu\nu} Q^{\mu \nu}.
\end{equation}

\section{\label{sec:renorm}Renormalization Factors}

In this section we complete the explicit one-loop renormalizability
calculuation.  We begin by taking logarithms of the expression
\eqref{expldet3.1} and substituting for the resulting divergences
using \eqref{YMcont1.1}, \eqref{fermcont3.1}, and
\eqref{ghostcont3.1}.  To treat the divergences arising from the pure
Yang-Mills term, we proceed by noting that $k_F$ can be treated as the
sum of selfdual and anti-selfdual parts \cite{cminst1}.
Noting that the selfdual contribution is trace-free,
we match the structure of the Lagrangian to the structure of the
corresponding singularity in the expression of the divergence
\eqref{YMcont1.1}.  The sum of corresponding terms is given by
\beq
{\mathcal L}_0 + \delta {\mathcal L} =
-\frac{1}{4g^2}\left(1-\frac{6g^2}{(4\pi)^2} \Gamma(2-\frac{d}{2})
 \right) (k_F)_{\mu \nu \alpha \beta} F^{\mu\nu}F^{\alpha\beta}~.
\eeq
Rescaling the bare parameters $g$ and $k_F$ via
\begin{eqnarray}
g_b & = & Z_g g_r \\
(k_F)_b & = & Z_{k_F} (k_F)_r~,
\end{eqnarray}
leads to
\begin{equation}\label{rescal3.3}
\frac{(k_F)_r}{g_r^2} = \frac{Z_g^2}{Z_{k_F}} \frac{(k_F)_b}{g_b^2}~.
\end{equation}
The calculation of $Z_g$ produces the same result as the standard
calculation for renormalizability of standard Yang-Mills
\begin{equation}\label{zkg1.1}
Z_g = 1 -  \frac{g^2 }{(4\pi)^2}\Gamma(2-\frac{d}{2})(\frac {11}{6} C_2(G) - \frac{2}{3} n_f C(r))~,
\end{equation}
where $n_f$ is the number of fermion species assumed to all be in representation $r$.
When $k_F$ is selfdual (that is, when $(k_F)_{\mu\nu\alpha\beta} =
\frac{1}{4}\epsilon_{\mu\nu\lambda\kappa}
(k_F)^{\lambda\kappa\rho\sigma} \epsilon_{\rho\sigma\alpha\beta},$ see
\cite{cminst1}).  This leads immediately to the scaling for $k_F:$
\begin{equation}\label{zkf1.1}
Z_{k_F} = 1 +  \frac{g^2}{(4\pi)^2}\Gamma(2-\frac{d}{2}) (\frac 7 3 C_2(G) + \frac 4 3 n_f C(r))~,
\end{equation}
which coincides with the trace-free result given in \cite{cm1}
in the absence of fermions.

For the anti-selfdual contribution, we note that
$k_F^{\mu\nu\alpha\beta} $ can be written in terms of
$\Lambda^{\mu\nu} = \frac12 k_{F \alpha}^{~~~\mu \alpha \nu}$ \cite{cminst1}:
\beq
k_F^{\mu\nu\alpha\beta} = \Lambda^{[\mu[\alpha }g^{\nu]\beta]}.
\eeq
In the absence of fermions and ghosts, term matching leads to the
expression
\beq
{\mathcal L}_{0} + \delta{\mathcal L} =
 -\frac{1}{g^2}\left(1-\frac{11}{3}\frac{C_2(G) g^2}{(4\pi)^2}
 \Gamma(2-\frac{d}{2})  \right) \Lambda^{\mu\nu} Q^{\mu\nu}~,
\eeq
where $Q^{\mu\nu}$ is as given in \eqref{qmunu}.  Given the rescaling
of $g$ in Eq.\eqref{zkg1.1}, we see that in this case $\Lambda$ (and hence
$k_F$) is unaffected by renormalization due to the pure Yang-Mills sector.
Only the fermions and ghosts contribute.

Adding a fermion and the ghosts we use \eqref{fermcont3.1} and
\eqref{ghostcont3.1} and match terms to obtain
\bea
{\mathcal L}_{0} + \delta{\mathcal L} = - \frac{1}{g_r^2}\left(1 - \frac 4 3 \frac {g^2} {(4 \pi)^2}
\Gamma(2 - \frac d 2) n_f C(r)\right)\nonumber  \\ \otimes
\left[\Lambda^{\mu\nu} +\frac{1}{6}\frac{ g^2}{(4\pi)^2} \Gamma(2-\frac{d}{2})
 \left( C(r) c_{\mu\nu}+ C_2(G) C_{\mu\nu}\right)\right]  Q^{\mu\nu}~, \nonumber \\
\eea
with $n_f = 1$ for one fermion.
Defining the renormalized $\La$ parameter using
\beq
{\mathcal L}_{0} + \delta{\mathcal L}  =  -\frac{1}{g_r^2}
\Lambda^{\mu\nu}_r Q_{\mu\nu}
\eeq
and defining
\beq \label{zkf1.2}
\Lambda_b^{\mu\nu}  =  (Z_{\Lambda})^{\mu\nu}_{\alpha \beta}
\Lambda_r^{\alpha\beta}
\eeq
yields the relationship
\bea
& & (Z_{\Lambda})^{\mu\nu}_{\alpha \beta}\left[ \La_b^{\al\be} + 
\frac 1 6 \frac {g^2} {(4 \pi)^2}\Gamma(2 - \frac d 2)\left( C(r) c^{\al\be}+ C_2(G) C^{\al\be}\right)\right] 
\nonumber \\
 & & = Z_S \La_b^{\mn}~,
\eea
with
\beq
Z_S = \left(1 + \frac 4 3 \frac {g^2} {(4 \pi)^2}
\Gamma(2 - \frac d 2) n_f C(r)\right)~.
\eeq
Finally, incorporating terms for arbitrary fermions is
straightforward as the contributions simply add.
This is accomplished in the above formula by letting $n_f$ be
arbitrary and making the replacement $c \rightarrow \sum_f c_f$ as
a sum over fermion species.

The CPT-odd terms $k_{AF}$ contain no divergent contributions, however,
they still need to be renormalized as the combination $k_{AF} / g^2$ appears
in the classical action.
This means that $Z_{k_{AF}} = Z_g^2$ at one-loop, in agreement
with the result found in \cite{cm1}.

\section{\label{sec:beta}Beta Functions}

Tacitly assuming for the moment that our renormalization prescription
can be extended to all orders,
the renormalization constants $Z_{k_F}$ and $Z_{k_{AF}}$ can be used to deduce
the one-loop beta functions for these parameters.
Following the developments presented in \cite{klp1}, use is made of
\beq
\beta_{x_j} = \lim_{\ep \rightarrow 0}\left[ - \rho_{x_j} a_1^j
+ \sum_{k=1}^N \rho_{x_k} x_k {\partial a_1^j \over \partial x_k}\right]~,
\eeq
where $x_j$ represents an arbitrary running coupling in the theory,
the parameters $\rho_{x_j}$ are determined by comparing the mass
dimension of the renormalized parameters to the bare parameters in $d
= 4 - 2 \ep$ dimensions, and the $a_1^j$ represent the first order
divergent contribution to the rescaling factor associated to the
variable $x_j.$  In more detail, writing
\beq
x_{jb} = \mu^{\rho_{x_j}\ep}Z_{x_j}x_j~.
\eeq
gives the values
\beq
\rho_g = 1~, \quad \rho_{k_F} = \rho_{k_{AF}} = 0~,
\eeq
and the $a_1^j$ are defined by the expansion
\begin{eqnarray}
Z_{x_j}x_j & = & x_j + \sum_{n=1}^\infty \frac{a_n^j}{\epsilon^n}~.
\end{eqnarray}
As in the QED case \cite{klp1}, the coupling $g$ completely controls the running of the
Lorentz-violating parameters.
The resulting beta function for $g$ is given by
\beq
\beta_g = -{g^3 \over(4 \pi)^2}\left(\frac {11} 3 C_2(G) - \frac 4 3 n_f C(r)\right)~,
\eeq
the same as the conventional case.  The beta function corresponding to
$k_{AF}$ is
\beq
\beta_{k_{AF}} = -{g^2 \over (4 \pi)^2}\left(\frac {22} 3 C_2(G) - \frac 8 3 n_f C(r) \right)k_{AF}~.
\eeq
where the Lorentz indices have been suppressed for simplicity.
The selfdual part of $k_F$ has the beta function
\begin{equation}
\beta_{k_F} = {g^2 \over (4 \pi)^2} \left( \frac {14} 3 C_2(G) + \frac 8 3 n_f C(r) \right)k_F.
\end{equation}
The anti-selfdual contributions coupled to the fermions and ghosts give
\begin{equation}
\beta_{\Lambda^{\mu\nu}} = - \frac{1}{3}{g^2 \over (4\pi)^2}
\left(C(r)\left[\sum_f c_f^{\mu\nu} - 8 n_f \La^{\mn}\right] + C_2(G)C^{\mu\nu}\right).
\end{equation}
Note that special values $c$ and $\La$ can lead to cancelation
in the beta function.

\section{\label{sec:brst}BRST Symmetry}

Due to the preservation of gauge invariance in the perturbed theory, the Lorentz-violating
action satisfies a standard Becchi-Rouet-Stora-Tyutin (BRST)\cite{brst} symmetry provided that there is
no explicit ghost violation introduced.
The ghost violation terms are not invariant under a standard BRST transformation,
but a specific form for the gauge fixing term can be chosen to maintain invariance.
However, this introduces additional violation into the photon propagator which can
be absorbed by a better choice of gauge.
This means that explicit Lorentz violation in the ghost sector alone can violate
the gauge symmetry as well.

This symmetry should ensure that the multiplicative renormalization will be consistent
to all orders by fixing the ratios of the relevant counter-terms in the renormalized
lagrangian.
This implies that all coupling constants ($g$, $k_F$, and $k_{AF}$) are universal
as $g$ is in the conventional case.
An explicit proof of this fact to all orders is beyond the scope of the present paper.
This fact agrees with the explicit one loop calculations performed in this paper.

\section{\label{sec:sum}Summary}

The functional determinant technique is particularly well suited to Lorentz
violation loop calculations as the traces conveniently preserve both the
observer Lorentz invariance and the gauge invariance throughout the calculation.
Sums over special subclasses of diagrams are required to maintain
a similar invariance using the diagrammatic approach \cite{cm1}.
In addition to the ease of organization of the calculation, the present approach is
well suited to exploring renormalization in more complicated versions of the
Lorentz-violating standard model.

New results of this paper include the contribution of the trace components of $k_f$,
terms that were neglected in the previous paper \cite{cm1}.
These terms in fact renormalize differently than the trace-free $k_f$ components
indicating their fundamentally different properties.
Additional new results include the incorporation of fermions and explicit ghost
violation.  Both effects are accommodated by a renormalization of the trace
components of $k_f$, while the trace-free components are unaffected.
These additional results give the entire one-loop gluon sector effective action in
Lorentz-violating QCD and demonstrate renormalizability at this order.

In addition, explicit BRST symmetry is present in the full Lorentz violating theory
(without explicit ghost violation) and should be crucial in eventually establishing
full renormalizability of the theory.

\begin{acknowledgments}
We wish to acknowledge the support of New College of Florida's faculty development
funds that contributed to the successful completion of this project.
\end{acknowledgments}

%\newpage %Just because of unusual number of tables stacked at end
\bibliography{apsfermrenorm}% Produces the bibliography via BibTeX.

\end{document}